\documentclass[submission, Phys]{SciPost_better_arXiv}

\hypersetup{
    colorlinks,
    linkcolor={red!50!black},
    citecolor={blue!50!black},
    urlcolor={blue!50!black}
}

\usepackage{pdfcomment}
\usepackage{sansmath} 

\DeclareSymbolFont{usualmathcal}{OMS}{cmsy}{m}{n}
\DeclareSymbolFontAlphabet{\mathcal}{usualmathcal}

\usepackage{booktabs,tabularx}
\usepackage{xspace}
\usepackage{subcaption}

\usepackage{enumitem}
\usepackage{amssymb}

\usepackage[utf8]{inputenc} 
\usepackage[compat=1.0.0]{tikz-feynman}
\usepackage{scalerel}
\usepackage{layouts}

\usepackage{tikz-layers}
\usepackage{lineno}
\usepackage{printlen}
\usepackage{wasysym}
\usepackage{grffile}

\newcommand{\email}[1]{\href{mailto:#1}{#1}}

\newcommand{\xB}{\ensuremath{x_{\mathrm{B}}}\xspace}
\newcommand{\ydis}{\ensuremath{y_{\mathrm{DIS}}}}
\newcommand\FASTJET{{\tt Fastjet}\xspace}
\newcommand\PYTHIA{{\tt Pythia}\xspace}
\newcommand\PYTHIAn{{\tt Pythia~\!\!8.3}\xspace}

\definecolor{darkgreen}{rgb}{0,0.4,0}
\definecolor{grey}{rgb}{0.5,0.5,0.5}
\definecolor{orange}{rgb}{0.9,0.5,0.0}
\definecolor{lightblue}{rgb}{0.0,0.5,1.0}
\definecolor{darkspringgreen}{rgb}{0.09, 0.45, 0.27}

\begin{document}
\begin{flushright}
CERN-TH-2026-032, 
Nikhef 2026-005, OUTP-26-02P
\end{flushright}
\begin{center}{\Large \textbf{
      A generalised-$k_t$ jet
algorithm for Deep Inelastic Scattering
\\ }}\end{center}

\begin{center}
  Melissa van Beekveld\textsuperscript{1},
  Silvia Ferrario Ravasio\textsuperscript{2},
  Alexander Karlberg\textsuperscript{3,4},
  Darcy Peake\textsuperscript{5}
\end{center}

\begin{center}
  {\small
{\bf 1} Nikhef, Theory Group, Science Park 105, 1098 XG, Amsterdam, The Netherlands \\
{\bf 2} Dipartimento di Fisica, Università di Torino, and INFN, Sezione di Torino, Via P. Giuria 1, I-10125 Torino, Italy \\
{\bf 3} Rudolf Peierls Centre for Theoretical Physics, Clarendon Laboratory, Parks Road, University of Oxford, Oxford OX1 3PU, UK \\
{\bf 4} CERN, Theoretical Physics Department, CH-1211 Geneva 23, Switzerland \\
{\bf 5} Department of Physics and Astronomy, University of Sussex, Sussex House, Brighton, BN1 9RH, UK \\[0.3cm]
  {\small \sf \email{mbeekvel@nikhef.nl},
    \email{silvia.ferrarioravasio@unito.it},
    \email{alexander.karlberg@cern.ch},
    \email{n.peake@sussex.ac.uk}
  }}
\end{center}

\section*{\color{scipostdeepblue}{Abstract}}
{\bf
We introduce an inclusive generalised-$k_t$ jet algorithm for Deep
Inelastic Scattering, defined in the Breit frame and implemented in
\texttt{fjcontrib}.  The family of algorithms is governed by the usual
parameter $p$, which controls the transverse-momentum dependence of
the algorithm, as well as by a jet radius parameter $R$. The
angular-ordered ($p=0$) version of the algorithm was already presented
by some of us, and can be used to formulate observables with simple
all-order structures.
In this article we investigate phenomenological applications of the
algorithms related to the identification of the jet associated with
the struck quark, and assess their sensitivity to non-perturbative
effects, such as hadronisation.
We also perform comparisons with the recent Centauro algorithm.
\begin{center}
  The code is available from
  \url{https://fastjet.fr/contrib/} as the plugin \href{https://repo.hepforge.org/source/fastjetsvn/browse/contrib/contribs/DISGenkt}{\texttt{DISGenkt}}.
\end{center}

}
\vspace{\baselineskip}
\noindent\textcolor{white!90!black}{%
\fbox{\parbox{0.975\linewidth}{%
\textcolor{white!40!black}{\begin{tabular}{lr}%
  \begin{minipage}{0.6\textwidth}%
    {\small Copyright attribution to authors. \newline
    This work is a submission to SciPost Phys. Comm. Rep. \newline
    License information to appear upon publication. \newline
    Publication information to appear upon publication.}
  \end{minipage} & \begin{minipage}{0.4\textwidth}
    {\small Received Date \newline Accepted Date \newline Published Date}%
  \end{minipage}
\end{tabular}}
}}}
\newpage

\vspace{10pt}
\noindent\rule{\textwidth}{1pt}
\tableofcontents\thispagestyle{fancy}
\noindent\rule{\textwidth}{1pt}
\vspace{10pt}

\section{Introduction}

Lepton-hadron colliders such as the Hadron Elektron
Ringanlage (HERA) and the forthcoming Electron--Ion Collider (EIC) are
unique laboratories for the clean investigation of the inner structure
of the nucleons and the nuclei that they form.
Although the HERA collider has not been operational since 2007, access
to previously recorded HERA data has enabled recent re-analyses taking
advantage of modern jet algorithms and event-shape observables in Deep
Inelastic Scattering (DIS)~\cite{ZEUS:2023zie,H1:2023fzk,ZEUS:2024mhu,H1:2024nde,H1:2024aze,H1:2024pvu}.
In the theory community this has spurred a renewed interest in
formulating jet
algorithms~\cite{Arratia:2020ssx,vanBeekveld:2023chs,Caucal:2024vbv},
and in performing resummed calculations of event shapes for
DIS~\cite{Makris:2021drz,Knobbe:2023ehi,Fang:2024auf,Cao:2024ota,Ee:2025scz,vanBeekveld:2025zjh}, the latter 
building upon earlier work (see e.g.\ Refs.~\cite{Antonelli:1999kx, Dasgupta:2001eq, Dasgupta:2003iq, Kang:2013nha, Kang:2013lga, Chu:2022jgs}).
Such modern QCD tools will be instrumental for the EIC,\footnote{And the LHeC~\cite{LHeC:2020van} should it be approved.} currently under construction at
Brookhaven National Laboratory, whose operation is scheduled to start
in the early 2030s.  The EIC will provide new opportunities to study
the structure of QCD, including the spin structure of the proton, the
dynamics of confinement, and gluon
saturation~\cite{Accardi:2012qut,AbdulKhalek:2021gbh}.
Since jet production is the dominant process at hadron-lepton
colliders, jet reconstruction will play an essential part in many
(re)-analyses for both the EIC and HERA.

Most developments in jet algorithms over the past decades have been
driven by hadron collider applications, in particular at the LHC,
where the anti-$k_t$ algorithm~\cite{Cacciari:2008gp} has become
the standard choice. Jet algorithms for DIS were developed around the
same time as those for hadronic collisions, but unlike for hadron
collisions, no single algorithm dominates the landscape.
The exclusive (spherically-invariant) $k_t$ algorithm was introduced
in Ref.~\cite{Catani:1992zp} extending the Durham algorithm for
$e^+e^-$ collisions~\cite{Catani:1991hj,Brown:1991hx} to DIS by
treating the proton remnant as a particle of infinite momentum. An
angular-ordered version of this algorithm was later proposed in
Ref.~\cite{Wobisch:1998wt}. It modifies the $e^+e^-$ Cambridge
algorithm~\cite{Dokshitzer:1997in} to DIS again by treating the proton
remnant as a particle of infinite momentum. A longitudinally invariant
(LI) $k_t$ algorithm was introduced in
Ref.~\cite{Ellis:1993tq,Catani:1993hr}, and was extended to an
angular-ordered version (the Aachen algorithm) in
Ref.~\cite{Wobisch:1998wt}. The latter two algorithms are identical to
the hadron-collider versions, except that the clustering takes place
in the Breit frame as is the case for all the DIS jet algorithms
listed here (cf.\ Sec.~\ref{sec:Kinematics} for details on the Breit
frame). More recently, the Centauro algorithm~\cite{Arratia:2020ssx}
was proposed, combining features of longitudinally-invariant $k_t$
algorithms with those of spherically-invariant algorithms. Likewise a
set of jet definitions suitable for small-\xB DIS were proposed in
Ref.~\cite{Caucal:2024vbv}. 
Most of the historical algorithms, including a modified version of the
JADE algorithm~\cite{JADE:1986kta,JADE:1988xlj}, have been studied
extensively at HERA and Fermilab's E665
experiment~\cite{E665:1992xqj,E665:1993vlk,H1:1994lps,ZEUS:1994jfw,ZEUS:1995tgg,H1:1995tux,H1:1998bvm,H1:1998rpm,H1:1998cuj,H1:2000bqr,H1:2002qhb,ZEUS:2002nms,ZEUS:2005iex,ZEUS:2006xvn,H1:2007xjj,H1:2009pqp,H1:2010mgp,H1:2014cbm},
highlighting the importance of jet finding algorithms in DIS. The
recent Centauro algorithm has also recently been used by the H1
collaboration to perform the first measurement of groomed event
shapes~\cite{H1:2024pvu} in DIS.

Unlike at the LHC, where one of the main requirements of a good jet
algorithm is the insensitivity to pile-up and underlying event, in
DIS, where these effects are suppressed, a good jet algorithm is one
that preserves a meaningful clustering sequence and allows soft
radiation to influence jet boundaries while respecting the natural
hemisphere structure of the Breit frame. For this reason it is natural
to consider $k_t$ and Cambridge/Aachen~(C/A) type algorithms in DIS,
in addition to anti-$k_t$. A C/A-based DIS algorithm designed for the
study of Lund variables~\cite{Dreyer:2018nbf,vanBeekveld:2025zjh} and
parton shower logarithmic accuracy was introduced in
Ref.~\cite{vanBeekveld:2023chs}, and forms the starting point for an
inclusive generalised-$k_t$ algorithm adapted to DIS,
which we present and study in more detail in the present paper. The
formulation of the jet algorithm itself follows the structure of the
generalised-$k_t$ algorithm used in $e^+e^-$ collisions as implemented
and introduced in \FASTJET~\cite{Cacciari:2011ma}, but is formulated
in the Breit frame and uses a beam distance that is sensitive to the
angular separation between the proton and the particles taking part in
the clustering. We also discuss a prescription to select the jet that
is most likely to contain the struck quark based on the lightcone
fraction that the jet is carrying. This allows for a clean separation
between \emph{target} and \emph{current} hemisphere jets. Finally, we
study the robustness of the new jet algorithm against hadronisation
corrections.

The structure of this paper is as follows.
In Section~\ref{sec:Kinematics} we introduce the kinematics relevant
to DIS and define the Breit frame.
In Section~\ref{sec:oldDISalgo} we summarise historical DIS jet
algorithms and the recently proposed Centauro one, while in
Section~\ref{sec:ourDISalgo} we present our new inclusive
generalised-$k_t$ jet algorithm.
In Section~\ref{sec:pheno} we examine the behaviour of the new
algorithms and we provide a comparison with the Centauro one.
Finally, we present our conclusions in Section~\ref{sec:concl}.

\section{DIS kinematics and the Breit Frame}
\label{sec:Kinematics}
The DIS process is defined by the partonic scattering process
\begin{equation}
\label{eq:DIS-Process}
    \ell_i(k) + h(P) \rightarrow \ell'_f(k') + X\,,
\end{equation}
where $P^\mu$ is the momentum of the incoming nucleon $h$,
$k^\mu$ is the momentum of the lepton beam $\ell_i$,
$k'^\mu$ is the momentum of the outgoing lepton $\ell'_f$,
and $X$ denotes the hadronic final state. 
The standard DIS invariants are
\begin{equation}
\label{eq:DIS-invariants}
    Q^2 = -q^2\,, \qquad
    \xB = \frac{Q^2}{2 P\!\cdot q}\,, \qquad
    \ydis = \frac{P\!\cdot q}{P\!\cdot k}\,,
\end{equation}
with $q^\mu = k^\mu - k'^\mu$ the momentum of the exchanged virtual photon.
At the lowest perturbative order, the hadronic final state consists of a single outgoing parton,
\begin{equation}
\label{eq:DIS-LO}
    \ell_i(k) + p_{\mathrm{in}} \rightarrow \ell_f(k') + p_{\mathrm{out}}\,,
\end{equation}
where $p_{\mathrm{in}}^\mu = \xB P^\mu$ and $p_{\mathrm{out}}^\mu = q^\mu
+p_{\mathrm{in}}^\mu$ are the momenta of the incoming and outgoing
partons, respectively.  The \emph{Breit frame} plays a central role in many DIS jet
algorithms~\cite{Webber:1993bm}.%
\footnote{The frame was
	first introduced in Ref.~\cite{Breit:1929zz} in the context of
	relativistic scattering kinematics. The term `Breit frame' was
	first coined by Ernst, Sachs, and Wali in
	Refs.~\cite{Ernst:1960zza,Sachs:1962zzc}. For an explicit
	implementation we refer the reader to Appendix 7.11 in
	Ref.~\cite{Devenish:2004pb}.} 
It is defined by
\begin{equation}
  2\xB \vec{P} + \vec{q} = 0\,,
\end{equation}
such that the virtual photon carries no energy component and is aligned along the $z$-axis, i.e.\
\begin{equation}
\label{eq:DIS-photon}
    q^\mu = (0,0,0,-Q)\,,
\end{equation}
and is therefore purely space-like.  
In this frame, the incoming and outgoing parton momenta take the simple form
\begin{equation}
\begin{aligned}
\label{eq:DIS-partons}
    p_{\mathrm{in}} ^\mu &= \frac{Q}{2}(1,0,0,+1)\,, \quad
    p_{\mathrm{out}}^{\mu} &= \frac{Q}{2}(1,0,0,-1)\,.
\end{aligned}
\end{equation}
Thus, the proton and virtual photon collide
head-on  in the Breit frame.
Within this frame, the proton remnant appears in the region at
positive rapidity (generally refered to as the \emph{target hemisphere}), while the struck
quark lies in the negative rapidity region (the \emph{current
hemisphere}).
In practice, the jet clustering in this frame can lead to radiation
from the proton remnant being absorbed into the struck-quark jet,
making the separation between the two hemispheres an important feature
of DIS jet algorithms.

\section{Existing DIS jet algorithms}
\label{sec:oldDISalgo}
Before presenting a generalised-$k_t$ algorithm for DIS in
Section~\ref{sec:ourDISalgo}, it is instructive to first discuss
existing jet algorithms that have historically been used for DIS,
focusing on sequential-recombination jet algorithms.
The overview we present here serves to explain how jet algorithms are formulated in general,
but will also allow us to highlight similarities and differences
between existing jet algorithms and the new generalised $k_t$ jet algorithm for DIS.
We formulate the jet algorithms in the Breit frame, but note that
those algorithms that are longitudinally boost invariant, such as
Centauro and some of the inclusive $k_t$-algorithms
(cf.\ Secs.~\ref{sec:gen-kt} and \ref{sec:centauro}), can also be
formulated in a frame that is connected to the Breit frame through a
longitudinal boost.\footnote{We note that the lab frame in DIS is
\emph{not} connected to the Breit frame through only a longitudinal
boost, but requires a rotation as well.}

\subsection{Exclusive jet algorithms for  $e^+e^-$ collisions}
Since some of the existing DIS jet algorithms are adapted from their $e^+e^-$ counterparts, and in fact use  $e^+e^-$ clustering explicitly in a re-clustering step, we start by reviewing the $e^+e^-$ variants before turning to DIS.
\subsubsection{The Durham jet algorithm}
\label{sec:dum-ee}
The Durham algorithm was introduced in Ref.~\cite{Catani:1991hj,Brown:1991hx}. 
Given a list of objects undergoing the clustering procedure, which are conventionally called pseudojets, one introduces a dimensionless resolution parameter $y_{ij}$ that reads
\begin{align}
	y_{i j} = \frac{\min (E_i^{2}, E_j^{2})}{s} 2(1 - \cos \theta_{i j})\,,
\end{align}
with $E_i, E_j$ the energies of the two pseudojets, $\theta_{ij}$ the angle between them, and $\sqrt{s}$ the center-of-mass energy of the collision.
One also introduces a parameter $y_{\rm cut} \leq 1$.
The algorithm then proceeds as follows. 
\begin{enumerate}
	\item \label{item1eekt}For every pair of pseudojets compute $y_{ij}$.
	\item The pair with the smallest $y_{ij}$ is clustered
          together if $y_{ij}<y_{\rm cut}$, meaning that a pseudojet
          with momentum $p_{ij} = p_i + p_j$ is inserted in the list
          of pseudojets, which replaces $i$ and $j$.
	  \footnote{This recombination is dubbed the $E$-scheme but other alternatives
	  exist~\cite{JADE:1986kta,JADE:1988xlj,Catani:1991hj}.}
	\item Repeat from \ref{item1eekt} until all $y_{ij} \geq
          y_{\rm cut}$ at which point the remaining pseudo-jets
          are identified as jets.
\end{enumerate}
After the clustering terminates, each particle in the event is either merged into a jet, or it remains as a single-particle jet, making it an exclusive algorithm. 
\subsubsection{The Cambridge jet algorithm}
\label{sec:cam-ee}
The Cambridge algorithm, introduced in
Ref.~\cite{Dokshitzer:1997in}, differs from the Durham algorithm by introducing, next to a merging condition, also a distance measure $d_{ij}$, which reads
\begin{align}
	d_{i j} = 2 (1 - \cos \theta_{i j})\,.
\end{align}
This choice was motivated by the aim of producing a clustering sequence that closely follows the coherent pattern of QCD radiation, where a possible merging of two partons that are close in angle is considered before the possible merge of two particles that are more widely separated. 
The merging of two particles is decided on the Durham distance measure $y_{ij}$.
Secondly, the clustering algorithm itself is modified with a `soft freezing' mechanism, which prevents the softer of two pseudojets to attract any additional pseudojets.
The Cambridge algorithm proceeds as follows. 
\begin{enumerate}
	\item \label{item1eeca} Compute $d_{ij}$ for all pairs and select the pair of pseudojets with the smallest $d_{ij}$.
	\item For this pair, if $y_{ij}<y_{\rm cut}$, delete $i$ and $j$ and introduce a new pseudojet with momentum $p_{ij} = p_i + p_j$. 
	If instead $y_{ij} \geq y_{\rm cut}$, the softer (i.e.\ less energetic)
	pseudojet is removed from the list and is promoted to a jet: this
	mechanism is called soft freezing. 
	\item Repeat from \ref{item1eeca} until there are no more pseudojets to be clustered. 
\end{enumerate}
The last pseudojet that remains in the list of pseudojets is also promoted to a jet.

\subsection{Exclusive jet algorithms for DIS collisions}
\subsubsection{Exclusive $k_t$-algorithm for DIS}
\label{sec:exclktalgo}
The exclusive $k_t$-algorithm for DIS was presented in
Ref.~\cite{Catani:1992zp}.
It is based on the equivalent $e^+e^-$ variant~\cite{Catani:1991hj} that we discussed above, but treats also the additional initial-state (and beam remnant) collinear singularities. 
The algorithm is formulated in the Breit frame. 
Instead of using the center-of-mass energy $\sqrt{s}$, one introduces a transverse-momentum-like scale $E_t$ satisfying $\Lambda \ll E_t\le
Q$ and defines a resolution variable $y_{ij}$ between two pseudojets
as
\begin{align}
\label{eq:disyij}
y_{i j} = \frac{\min (E_i^2, E_j^2)}{E_t^2} 2 (1 - \cos \theta_{i j})\,,
\end{align}
as well as a beam resolution parameter for each pseudojet
\begin{align}
	\label{eq:disyB}
  y_{iB} = \frac{ E_i^2}{E_t^2} 2(1 - \cos \theta_{i B})\,,
\end{align}
where $\theta_{ij}$ is the angular separation between two pseudojets,
$\theta_{iB}$ is the angle of pseudojet $i$ with respect to the beam, and
$E_{i}$ and $E_j$ are the energies of the corresponding pseudojets.
The $E_t$ variable determines how close to the beam direction (in the Breit frame) resulting jets are allowed to be formed, with the transverse momentum of the resulting jets satisfying $p_t > E_t$. 
The clustering procedure works in two steps.  
In the first step, a pre-clustering takes place, with the aim of separating particles into a \emph{beam} and a set of \emph{macro-jets}, and operates as follows.
\begin{enumerate}
  \item \label{exclkt1} For every pseudojet $i$ compute $y_{iB}$ and for every pair
    $i,j$ compute $y_{ij}$.
  \item Take the smallest value of all $y_{ij}$ and $y_{iB}$. If the
    smallest value is a $y_{ij}$ and $y_{ij}<1$, combine their momenta
    into a new pseudojet. If $y_{iB}<1$ is the smallest, include $i$
    into the beam jet, and remove it from the list of pseudojets. 
  \item Repeat the procedure from step \ref{exclkt1} for all
    pseudojets, until all distance measures are $y_{ij}, y_{iB}>1$. 
\end{enumerate}
The resulting objects that remain at this stage are the final-state macro-jets, and does not include the beam jet. 
In the second step of the algorithm we resolve the structure of the final-state macro-jets. 
The input for this second step consists of all the constituents of these jets. 
At this stage, a resolution parameter $y_{\rm cut} = Q_0^2/E_t^2 < 1$ is introduced, with $Q_0 < E_t$, and the algorithm proceeds by clustering the constituents using the $e^+e^-$ Durham algorithm (Sec.~\ref{sec:dum-ee}). 
The objects that remain at the end are the final-state jets. 
The full two-step procedure produces jets with relative transverse momenta $k_t$ in the range $\sqrt{y_{\rm cut}}E_t < k_t < E_t$, and with a Breit-frame transverse momentum $p_t > E_t$. 
Every particle in the event ends up either in the list of final-state jets or in the beam jet. 

\subsubsection{Cambridge algorithm for DIS}
\label{sec:camdis}
The Cambridge algorithm for DIS was introduced in
Ref.~\cite{Wobisch:1998wt}. It is an exclusive algorithm, inspired by the Cambridge
algorithm for $e^+e^-$-collisions~\cite{Dokshitzer:1997in }, and
very similar to the exclusive $k_t$-algorithm presented in Sec.~\ref{sec:exclktalgo}.
Like in $e^+e^-$, one decouples the merging step from the ordering step, now by introducing  two distance measures: $d_{ij}$ and $d_{iB}$, which read
\begin{align}
d_{i j} = 2 (1 - \cos \theta_{i j})\,,
\end{align}
and 
\begin{align}
d_{i B} = 2 (1 - \cos \theta_{i B})\,.
\end{align}
Like for the exclusive $k_t$-algorithm presented in Sec.~\ref{sec:exclktalgo}, also in this case the algorithm proceeds in two steps. 
The first step is
\begin{enumerate}
  \item \label{cam1} For every pseudojet $i$ compute $d_{iB}$ and for every pair
    $i,j$ compute $d_{ij}$.
  \item Take the smallest value of all $d_{ij}$ and $d_{iB}$.
    If the
    smallest value is a $d_{ij}$, combine their momenta
    into a new pseudojet if $y_{ij}<1$ otherwise call the softest (less energetic) a jet (i.e.\ apply the freezing mechanism). 
    If $d_{iB}$ is the smallest include $i$ as a beam jet, and remove
    it from the list of pseudojets if $y_{iB}<1$, otherwise promote it
    to a jet.
  \item Repeat the procedure from step \ref{cam1} for all
    pseudojets, until all resolution variables $y_{ij}, y_{iB}>1$,
    at which point we promote all the remaining pseudojets to jets and proceed with the second step. 
\end{enumerate}
In the second step of the algorithm we undo the clustering of all the
jets (excluding the beam jet), and re-cluster the constituents using
the $e^+e^-$ Cambridge algorithm using a resolution parameter
$y_{\text{cut}}$.

\subsection{Generalised inclusive $k_t$-algorithms for DIS}
\label{sec:gen-kt}
The jet algorithms for DIS discussed so far were all
\textbf{exclusive}. For an \textbf{inclusive} algorithm, when the
smallest distance is the one between the beam(s) and a pseudo-jet, the
pseudo-jet is promoted to a jet rather than being absorbed into a beam
remnant collection. The total number of jets is therefore an infrared
unsafe observable, because it includes jets that can be either very
soft or collinear with the beam. Only after introducing a suitable
resolution parameter (like a minimum transverse momentum with respect
to the beam) and removing jets that fall below this resolution
parameter, will it be infrared safe.

The inclusive longitudinally-invariant~(LI) $k_t$-algorithm was first
introduced in Ref.~\cite{Catani:1993hr} (see Ref.~\cite{Ellis:1993tq}
for the exclusive version) for hadron-hadron collisions. 
An angular-ordered variant, often referred to as the Aachen or Cambridge/Aachen algorithm,
was since introduced in Ref.~\cite{Wobisch:1998wt} and
then eventually the family of generalised inclusive $k_t$-algorithms
was introduced in Ref.~\cite{Cacciari:2008gp} along with the anti-$k_t$
algorithm.
The DIS variant is in this case identical to the hadron-hadron
variant, except that the clustering takes place in the Breit frame.
We define a set of distance measures $d_{ij}$ and $d_{iB}$, which are dimensionful for $p\neq 0$,  given by
\begin{align}
  d_{ij} = \min(p_{Ti}^{2p},p_{Tj}^{2p}) \frac{\Delta R_{ij}^2}{R^2}\,,\quad d_{iB} = p_{Ti}^{2p}\,,
\end{align}
where $\Delta R_{ij}^2 = (y_i -y_j)^2 + (\phi_i - \phi_j)^2$, with
$y_i$ the rapidity and $\phi_i$ the azimuthal angle of pseudojet
$i$. The procedure is then given by
\begin{enumerate}
\item Compute all the distances $d_{ij}$ and $d_{iB}$. If a $d_{ij}$
  is the smallest, cluster $i$ and $j$, if $d_{iB}$ is the smallest
  call $i$ a jet.
\item Repeat until no more pseudojets remain.
\end{enumerate}
We note that the $k_t$-algorithm corresponds to $p=1$,
C/A to $p=0$ and anti-$k_t$ to $p=-1$.

Note that one is not restricted to using distances that are natural for $pp$ collisions. 
Indeed, the inclusive algorithm can be made spherically invariant~(SI) instead of LI, which is more appropriate for $e^+e^-$ collisions~\cite{Cacciari:2011ma}, by
replacing the distance measures with
\begin{align}
\label{eq:eegen}
  d_{ij} = \min(E_{i}^{2p},E_{j}^{2p}) \frac{1-\cos\theta_{ij}}{f(R)}\,,\quad d_{iB} = E_{i}^{2p}\,,
\end{align}
with
\begin{equation}
  f(R) = \begin{cases}
    &1-\cos R \qquad 0< R < \pi \\
    &3+\cos R \qquad  \pi < R < 2\pi.
  \end{cases}
  \label{eq:fR}
\end{equation}
In this case, the beam distance $d_{iB}$ is not sensitive to the
angular separation between the pseudo-jet $i$ and the beam, but only
on the energy of the pseudo-jet.
This feature is not desirable in the context of $ep$ collisions,
which is an asymmetrical collision, where we need to distinguish
between a collinear and an anti-collinear direction, as there is only
one hadronic beam.

\subsection{Centauro algorithm}
\label{sec:centauro}
The Centauro algorithm~\cite{Arratia:2020ssx} is a longitudinally
invariant jet algorithm designed to capture jets close to the Born
configuration in the Breit frame. Longitudinally invariant $k_t$-type
algorithms such as those presented in Sec.~\ref{sec:gen-kt} cluster in the
$(y,\phi)$ plane and are unable to form a jet along the beam axis,
where $y \to -\infty$. Conversely, the spherically invariant
algorithms can cluster around the beam axis but do not preserve
longitudinal invariance.

The Centauro algorithm
is designed to address both of these aspects. It is built around
the variable $\bar{\eta}_i$ that behaves
differently in the two hemispheres:
\begin{itemize}
\item \textbf{Current hemisphere:}  
      $\bar{\eta}_i$ decreases as particles become closer in angle, allowing particles in this hemisphere to cluster.
\item \textbf{Target hemisphere:}  
      $\bar{\eta}_i$ becomes large and diverges. This prevents clustering in the proton beam region, similar to the behaviour of the anti-$k_t$-algorithm.
\end{itemize}
Working in the Breit frame, for every momentum $p_i^\mu$, one defines the variable $\bar{\eta}_i$ as
\begin{align}
\bar{\eta}_i = -2\frac{Q}{q\cdot n_{\rm out}} \frac{p_{T,i}}{p_i \cdot n_{\rm in}}\,,
\end{align}
with 
\begin{subequations}
  \begin{align}
  n_{\rm in}^\mu =& \frac{2}{Q}\xB P^\mu \\
  n_{\rm out}^\mu =& \frac{2}{Q} \left(q^\mu +\xB P^\mu \right)\\
  p_{T,i}^2 =& \frac{2 (p_i \cdot n_{\rm in})(p_i \cdot n_{\rm out})}{ n_{\rm in}\cdot n_{\rm out}},
\end{align}
\end{subequations}
meaning that $p_{T,i}$ is the transverse component of the momentum in the Breit frame (or any other frame that only differs by a longitudinal boost from the Breit frame).
From this, one constructs the inter-pseudojet distance measure as
\footnote{Note that we are reporting the definition used in the \texttt{fjcontrib} implementation, which is the one we have also used in this work.}
\begin{align}
\label{eq:cent-dij}
d_{ij} = \frac{(\bar{\eta}_i - \bar{\eta}_j)^2 + 2 \bar{\eta}_i \bar{\eta}_j (1- \cos \Delta \phi_{ij})}{R^2}\,.
\end{align}
The distance to the beam is always equal to $d_{iB} = 1$. 
One then clusters jets according to the procedure given in Sec.~\ref{sec:gen-kt}.

\section{A generalised-$k_t$ algorithm for DIS}
\label{sec:ourDISalgo}
\subsection{The algorithm}
We now introduce a generalised-$k_t$ algorithm for DIS, implemented as  a \textsc{FastJet} plugin in \texttt{fjcontrib}.
The algorithm shares the essential structure of its $e^+e^-$
counterpart, introduced at the end of Sec.~\ref{sec:gen-kt}, with the
key modification that the beam distance is now sensitive to the
angular separation with respect to the proton beam and the clustering
takes place in the Breit frame.
The angular-ordered variant of the algorithm ($p=0$) was already presented by some of us in Ref.~\cite{vanBeekveld:2023chs}, and used to define Lund Tree Event shapes with remarkably simple all-orders structure~\cite{vanBeekveld:2025zjh}.
Here we present the extension for $p\neq 0$, and include two parameters, $R$ and $y_{\rm cut}$ (to be detailed below), that control the behaviour of the algorithm. 
We first define the distances
\begin{subequations}
\begin{align}
	d_{ij}& = \min(E_{i}^{2p},E_{j}^{2p}) \frac{(1-\cos\theta_{ij})}{1-\cos R} \,,  \\
	d_{iB} &= E_{i}^{2p}(1-\cos\theta_{iB})\,, 
\end{align}
\end{subequations}
where $R < \pi$. 
The algorithm is then as follows.
\begin{enumerate}
	\item Compute all $d_{ij}$ and $d_{iB}$. If the smallest is a $d_{iB}$, call $i$ a jet, and remove it from the list of pseudojets. If the smallest is a $d_{ij}$, cluster $i$ and $j$. 
	\item Continue until there are no pseudojets left in the list. 
\end{enumerate}
The introduction of the distance $d_{iB}$ measuring both the energy
and the closeness in angle to the beam direction, allows one to
cluster jets in the current hemisphere while promoting remnant
hemisphere particles to jets early in the cluster sequence.
Note that the C/A variant ($p=0$) presented in
Ref.~\cite{vanBeekveld:2023chs} uses $R=\frac{\pi}{2}$.\footnote{It is also
possible to formulate an exclusive variant of our algorithm by
introducing a dimensionless quantity $y_{ij}$ and an associated
$y_{\mathrm{cut}}$. The $p=0$ zero variant will then share some
similarities with the algorithm in Sec.~\ref{sec:camdis} but without the
final re-clustering step. We do not explore this variant any further
in this paper. }
For $p=0$, the energy of a particle does not enter the distance measure, and particles are merged purely based on angular proximity. 
For $p=1$, i.e.\ the $k_t$ variant of the algorithm, soft particles typically lead to smaller $d_{ij}$ and therefore are clustered first.
On the contrary, for $p = -1$, i.e.\ the anti-$k_t$ variant, hard final-state particles will cluster first.

As in the standard inclusive (LI) algorithms for $pp$ collisions, some
reconstructed jets may originate entirely from initial-state radiation
or from soft large-angle emissions.
In $pp$ collisions, such jets are removed by imposing a
transverse-momentum (and a rapidity) cut.
This however cannot be applied to DIS, as the jet originating from the
fragmentation of the struck quark does not necessarily carry a large
transverse momentum (in fact, in the case where no additional
radiation is produced, this jet will carry zero transverse momentum).
To remove only soft jets or jets collinear to the beam, one can define
an ordering variable according to the (scaled) effective transverse
momentum
\begin{equation}
  y_{iB} = 2\frac{E_i^2 }{Q^2}(1-\cos\theta_{iB}),
\end{equation}
and discard jets with $y_{iB} < y_{\rm cut}$. 
Alternatively, one might use an ordering variable based on the light-cone
fraction of the original struck-quark momentum carried by the $i^{\rm
  th}$ jet, defined by
\begin{equation}
  z_{i} =  \frac{2 \xB\, p_i \cdot P}{Q^2} = 2\frac{E_i}{Q}(1-\cos\theta_{iB}),\label{eq:zi}
\end{equation}
where $P$ is the proton momentum. One may then discard jets with $z_i
< z_{\rm cut}$, with $0<z_{\rm cut}<1$. Below, we discuss how to use
$z_i$ to select what we will call the \emph{macrojet}.
Differences between Centauro (Sec.~\ref{sec:centauro}) and the
generalised-$k_t$ jet algorithm presented here will be explored in
Sec.~\ref{sec:pheno}.

\subsection{Identifying the macrojet}
\label{sec:macrojet}
\begin{figure}[t!]
\includegraphics[width=\textwidth,page=10]{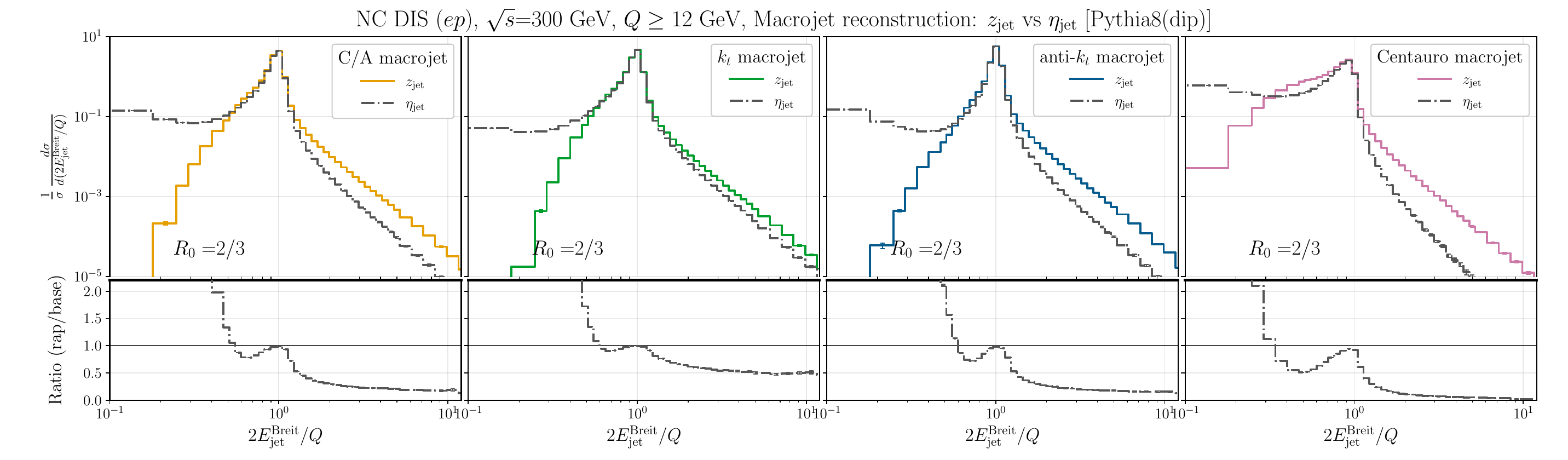}
\caption{Normalised Breit-frame energy distribution of the final-state
  macrojet for neutral-current DIS, $\sqrt{s}$=300~GeV using, from
  left to right, the inclusive version of the generalised $k_t$
  algorithms with $R_0 = 2/3$ and $p=0$ (C/A), $1$ ($k_t$), and $p=-1$
  (anti-$k_t$), followed by the Centauro jet algorithm with $R_0 =
  2/3$.
The final-state macrojet is identified as the one that maximises the light-cone component opposite in direction to the proton ($z_{\rm jet}$, solid line) or the Breit-frame pseudo-rapidity ($\eta_{\rm jet}$, dash-dotted lines).
Events are generated using the \PYTHIAn~\cite{Bierlich:2022pfr} default shower, with dipole recoil option~\cite{Cabouat:2017rzi}.
See Tab.~\ref{tab:input-params} for further settings.
 }
 \label{fig:jetEIR}
\end{figure}
This section describes how we identify the \emph{final-state
macrojet}, defined as the jet that is most likely to contain the 
fragments of the struck quark.
Our prescription is to select the jet that carries the largest
light-cone component $z_{\rm jet}$ along the direction of 
the original struck quark.
To make this precise, we perform a Sudakov decomposition of each
jet momentum (see also Sec.~\ref{sec:Kinematics} and Ref.~\cite{Makris:2021drz})
\begin{align}
	\label{eq:sudakov}
  p_{\rm jet}^\mu = p_{\rm jet}^- \,\xB\ P^\mu + z_{\rm jet} \, p_{\rm out}^\mu + k_\perp^\mu\,,
\end{align}
where $P^\mu$ is the proton momentum, and as before, $p_{\rm out}^\mu
= q^\mu + x_{\rm B} P^{\mu}$ is the original struck-quark
momentum. The coefficient $p_{\rm jet}^-$ is given by
\begin{equation}
  p^-_{\rm jet} = \frac{|\vec{k}_\perp|^2 + p_{\rm jet}^2}{Q^2 z_{\rm jet}} \, .
\end{equation}
The vector $k_\perp^\mu$ is a space-like vector orthogonal to
$P^\mu$ and $p_{\rm out}^\mu$.\footnote{Note that in Eq.~\eqref{eq:sudakov} we have adopted the notation of Ref.~\cite{Makris:2021drz} to label the lightcone component along the proton in the `$-$' direction.}
Maximising $z_{\rm jet}$ is then equivalent to selecting the jet whose
momentum $p_{\rm jet}^{\mu}$ has the largest dot product with
$P^\mu$. As discussed in Ref.~\cite{vanBeekveld:2023chs}, this
prescription is infrared safe.
A seemingly natural alternative is to identify the macrojet as the jet
with the most negative rapidity.
However, this prescription is not infrared safe, as illustrated in Fig.~\ref{fig:jetEIR}. 
There, one can observe that both for the new family
of jet algorithms introduced here and for the Centauro jet algorithm, identifying the
final-state macrojet by rapidity (dash-dotted lines) leads to a soft tail. 
This tail is produced by wrongfully identifying soft emissions, generated either from radiation off the
DIS hard-scattering process or from the beam remnant, as the final-state macrojet. 
This means that selecting the jet with the most negative rapidity is not infrared safe because an arbitrarily soft emission at sufficiently negative rapidity can alter the jet assignment discontinuously.
We note that for the $k_t$-algorithm the
soft tail is partially suppressed even when selecting the macrojet with the
most negative rapidity, but there is no exponential decrease in the tail.
Maximising the light-cone component in the direction of the 
original struck quark (solid curves) exponentially suppresses the tail for all
algorithms.

Beyond macrojet identification, the properties of the macrojet itself provide a useful way
to characterise the behaviour of the algorithm.
In particular, the macrojet offers a 
direct handle separating the remnant hemisphere from the current hemisphere.
We will explore this further in the next section.

We conclude this section by explicitly discussing the infrared and
collinear safety of the jet algorithm and the macrojet definition. In
the Born limit, the only final-state parton lies in the current
hemisphere and is promoted to a single jet with $z_{\rm jet}=1$, so
that the macrojet prescription reproduces the struck-quark
direction. Collinear splittings are clustered by the recombination
step before they can affect the resolved jet structure, and the
macrojet selection remains unchanged because $z_{\rm jet}$ is additive
under such splittings, preserving the total light-cone momentum of the
parent jet. Soft emissions either cluster with a nearby hard pseudojet
or form separate soft jets whose light-cone fraction vanishes in the
soft limit. They therefore cannot change the largest-$z_{\rm jet}$
assignment except in degenerate configurations of measure zero. The
resolved macrojet observables considered below are consequently
insensitive to arbitrarily soft emissions and collinear splittings.

\section{Phenomenological results}
\label{sec:pheno}
In this section we explore a few macrojet distributions obtained with
our inclusive jet algorithms. In what follows we will simply label the
$p=-1$ variant of the algorithm with anti-$k_t$, $p=0$ with C/A, and
$p=1$ with $k_t$. We compare them to the Centauro algorithm to
highlight differences and similarities.

DIS neutral-current events are produced using the
\PYTHIAn~\cite{Bierlich:2022pfr} event generator, with 
the native \PYTHIA shower with dipole local
recoil~\cite{Cabouat:2017rzi}. 
The analysis is performed with \textsc{Rivet~3.1.8}~\cite{Bierlich:2019rhm}. 
The parameters, unless otherwise stated, used for the results presented 
in the following sections are summarised in Table~\ref{tab:input-params}.
In Sec.~\ref{sec:all} we investigate the properties of all the reconstructed jets, while in Sec.~\ref{sec:obs-macrojet} and~\ref{sec:hadronisation} we focus on the final-state macrojet distributions.
\begin{table}[t]
\begin{center}
\begin{tabular}{c|c}
\hline\hline
Input Parameter & Value \\
\hline\hline
PDF & NNPDF40MC\_nlo\_as\_01180 \\
$\sqrt{s}$ & $300\,\mathrm{GeV}$\\
$Q_{\mathrm{min}}$ & $12\,\mathrm{GeV}$ \\
Hadronisation effects & On \\
Beam remnants & On \\
QED & Off \\
Tune & Monash~\cite{Skands:2014pea}\\
\hline\hline
\end{tabular}
\end{center}
\caption{Input parameters used for the numerical results presented below.}
\label{tab:input-params}
\end{table}

\subsection{Properties of the reconstructed jets}
\label{sec:all}
\begin{figure}[t!]
	\includegraphics[width=\textwidth,page=2]{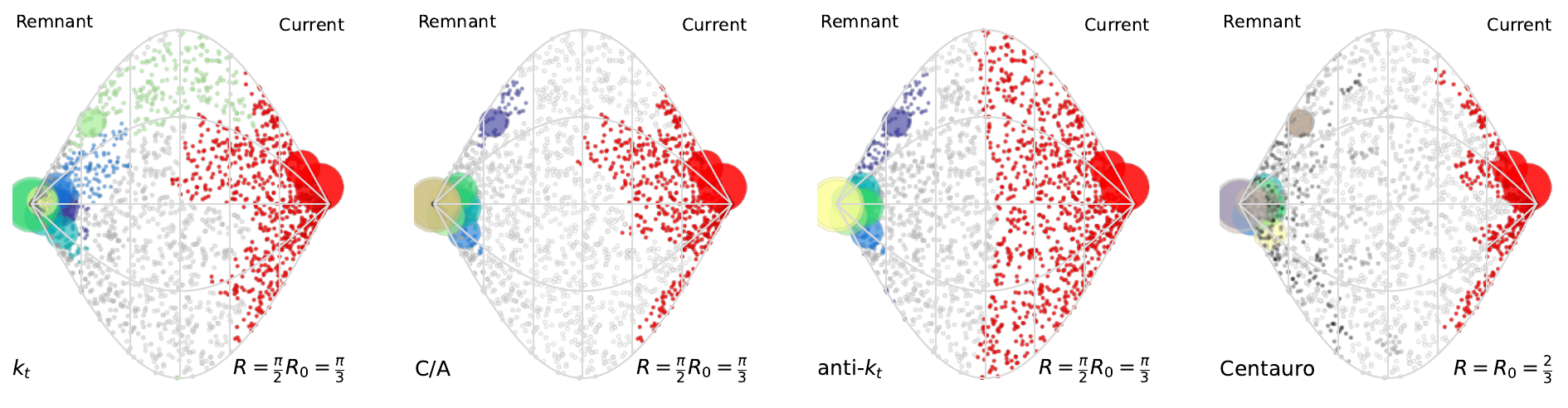}
	\caption{From left to right, we show the jet and jet boundaries for the $k_t$ ($p=1$), C/A ($p=0$), anti-$k_t$ ($p=-1$) and Centauro jet algorithms for a DIS event with $x_B = 0.2$ and $Q = 114$ GeV.
		Each particle is
		represented by a disc whose area is proportional to its energy,
		while its position corresponds to the direction of its momentum
		projected onto the unfolded sphere around the hard-scattering
		vertex. 
		In addition to the physical particles, we have added a sea of $2000$ ghosts~\cite{Cacciari:2008gn} to
		better illustrate how the shape of the jet boundaries. 
		The soft particles that are not part of any physical jet are coloured gray. 
		Vertical lines indicate constant $\theta$, and curved lines
		indicate constant $\phi$.
                The remnant hemisphere lies in the left half of the
                plot and the current hemisphere in the right half of
                the plot.
		All particles clustered into a given jet share
		the same color, and jets only made out of ghost particles are coloured gray. 
		Particles entering the final-state macro-jet are
		highlighted in red.
	}
	\label{fig:events}
\end{figure}

To better understand general features of the new jet algorithms and how they
compare to the Centauro jet algorithm, we show in
Fig.~\ref{fig:events} the $(\theta,\phi)$ distribution of particles
from a DIS event obtained after showering (but before hadronisation).
The location of the central point in each disc corresponds to the
particle's position in the $\theta$ and $\phi$ plane, while the size
of the disc illustrates the particle's energy, with
larger radii corresponding to higher relative energies.
Particles in the remnant hemisphere lie on the left-hand side of each
figure, while particles in the current hemisphere lie on the
right-hand side.
Particles entering the same jet share the same colour.
Those forming the final-state macrojet are coloured in red.
We have placed the event in a sea of 2000 additional ultra-soft
particles, dubbed ghosts in
Ref.~\cite{Cacciari:2008gn}, where the idea of illustrating the jet
boundaries was originally introduced.
We generate these ghosts with a fixed energy ($E = 10^{-6}\, \mathrm{GeV}$) and with a uniform 
distribution in $\theta$ and $\phi$. 
We stress here that the exact shape of the jet boundaries depends
to some extent on the exact distribution of the soft particles. 

Upon inspection of Fig.~\ref{fig:events}, one of the most striking visual features is that of the jet boundary produced by the anti-$k_t$ algorithm.
We see that the jet boundary for the final-state macrojet in this case follows the lines of constant $\theta$.
This is a direct consequence of the anti-$k_t$ clustering sequence, which always initiates clustering from the hardest particle outward, producing perfectly conical jets in the $\theta$,$\phi$ plane. 
This circular symmetry is not shared by the other algorithms: $k_t$, C/A and Centauro all produce jets with irregular, jagged boundaries.

Following Ref.~\cite{Arratia:2020ssx}, we adopt 
\begin{align}
\label{eq:R-comp}
R_{{\rm generalised}-k_t} = \frac{\pi}{2} \, R_{\rm centauro} \equiv  \frac{\pi}{2} \, R_0\,,
\end{align}
to compare the generalised-$k_t$ variants with Centauro. In the following phenomenological comparisons we quote radii in terms of $R_0$.
One other notable feature is that Centauro tends to produce smaller jets with similar-sized jet radius parameter $R_0$.
This is a consequence of a general property of the Centauro algorithm: it produces comparatively isolated, tight jets. 
This feature is even more pronounced in the remnant hemisphere, where
the size of the jet clusters is much bigger for all variants of the
generalised-$k_t$ jet algorithm than for Centauro, which exponentially
suppresses clustering of particles in the remnant hemisphere.

Note that for the $k_t$ and C/A variants of the generalised-$k_t$ algorithm, and for Centauro, the exact set of particles clustered into the \emph{macrojet} can change significantly with the choice of $R_0$. 
This occurs because these algorithms cluster particles based on relative softness or angular proximity, so in events with several particles carrying comparable energies, small changes in $R_0$ can cause different particles to be pulled into or out of the macrojet (as our definition of the macrojet relies on $z_{\rm jet}$). 
The anti-$k_t$ variant does not suffer from this, since it always clusters outward from the hardest particle, making the macrojet constituents much more stable against variations in $R_0$.

Despite its attractive properties, it is worth pointing out that depending on the choice of $R_0$, the anti-$k_t$ variant of our algorithm carries a specific feature. 
Precisely because it clusters outward from the hardest particle without regard for the hemisphere boundary, it can in principle absorb particles from the remnant hemisphere into the current-hemisphere macrojet. 
Empirically, values around $R_0\lesssim 2/3$ are found to significantly reduce contamination of the current-hemisphere macrojet by remnant radiation.

We end this subsection by studying general properties of all jets.
\begin{figure}[t!]
  \centering
  \includegraphics[width=0.45\textwidth,page=3]{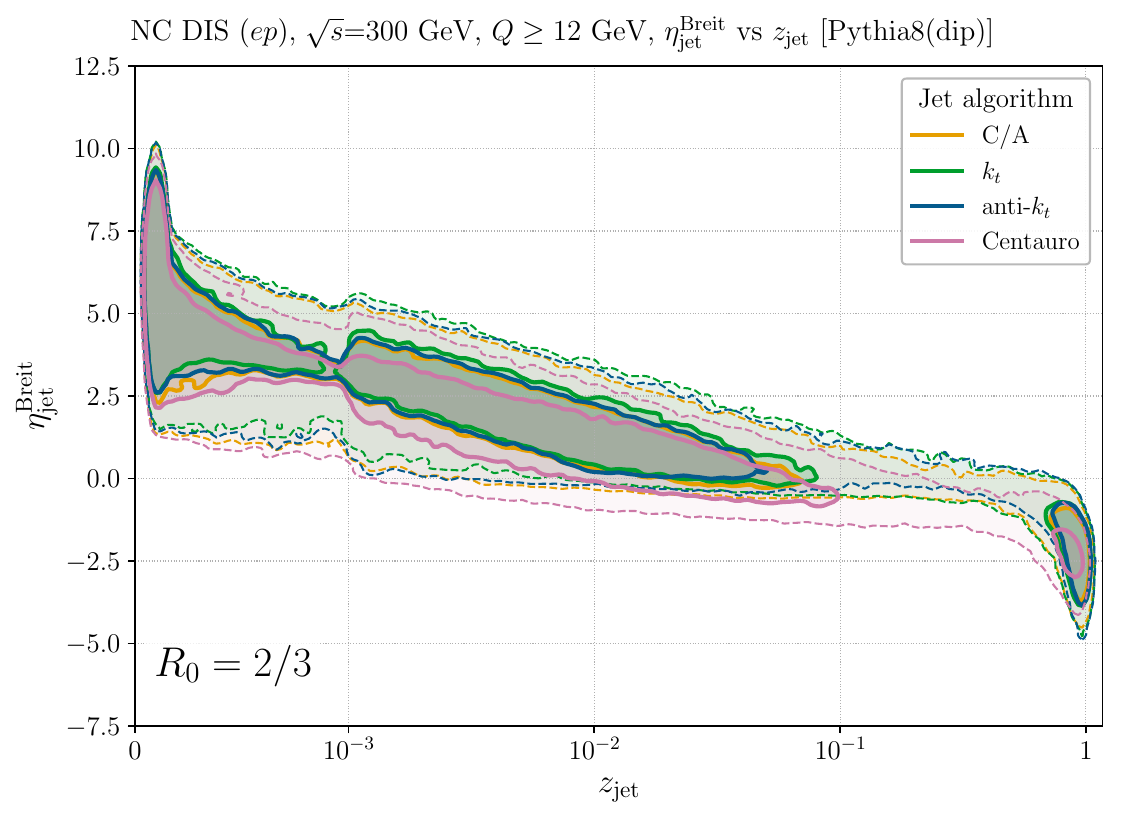}
  \includegraphics[width=0.45\textwidth,page=3]{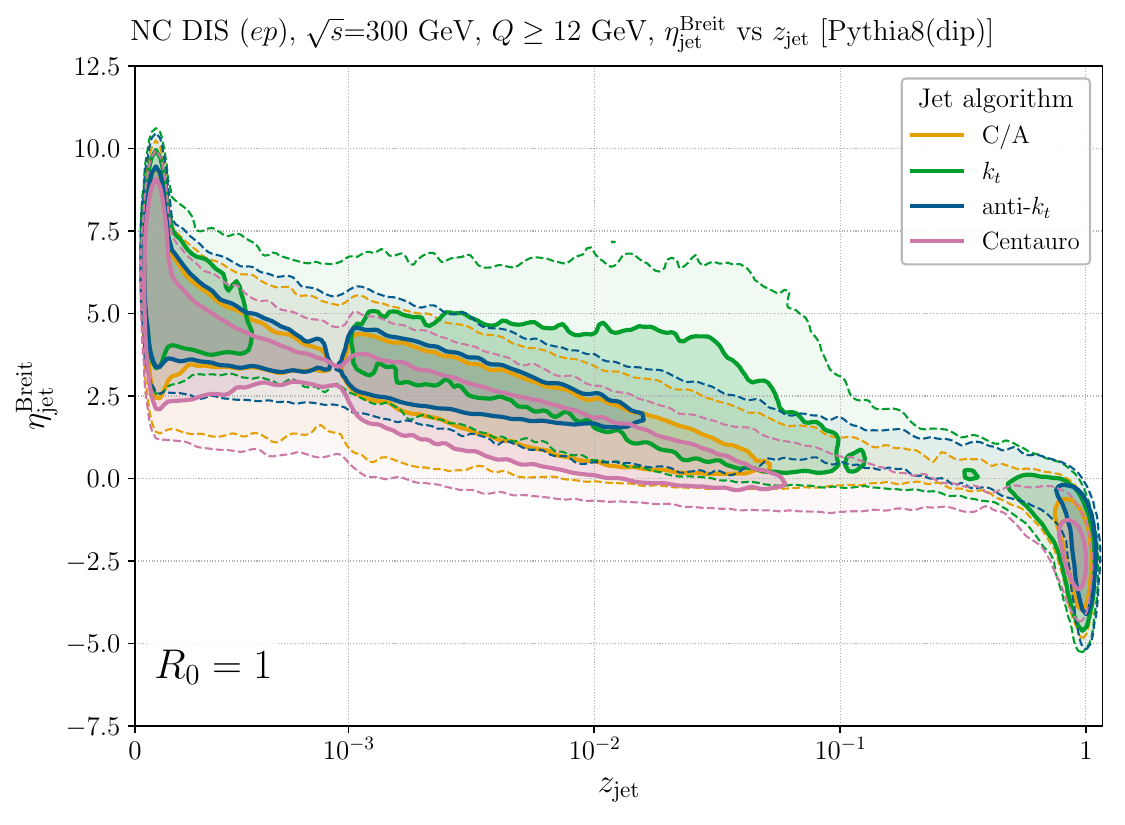}
\caption{The $z-\eta$-plane for all jets constructed with the Centauro
  (pink), C/A (orange), $k_t$ (green) and anti-$k_t$ (blue) algorithms
  for $R_0 = 2/3$ (left) and $R_0 = 1$ (right), using
  Eq.~\eqref{eq:R-comp}.  The solid (dashed) curves correspond to
  $68\%$ ($95\%$) of the jets, in terms of their density in the $z-\eta$-plane.}
  \label{fig:zeta}
\end{figure}
Fig.~\ref{fig:zeta} shows the $z-\eta$ plane for all final-state jets
constructed with the four algorithms considering $10^6$
events. 
Here, $z_{\rm jet}$, the  lightcone-momentum fraction carried by each of the final-state jets, is defined through Eq.~\eqref{eq:sudakov} as
\begin{equation}
	\label{eq:zjet}
	z_{\mathrm{jet}} = \frac{P \cdot p_{\mathrm{jet}}}{P \cdot q} = \frac{p_{\mathrm{jet}}^E - p_{\mathrm{jet}}^z}{|Q|}\,,
\end{equation} 
where the last equation holds in the Breit frame. We show contours
corresponding to areas that contain $68\%$ (solid) and $95\%$ (dashed)
of the total number of jets in the plot, for each of the four
considered jet algorithms, and for $R_0 = 2/3$ (left) and $R_0 = 1$
(right).

We observe that the jet algorithms behave relatively similar for
$R_0=2/3$ , where we observe two dense regions: one at small $z_{\rm
  jet}$ and (large and) positive pseudo-rapidity, and one at large
$z_{\rm jet}$ and negative rapidity. The large-$z_{\rm jet}$ region
corresponds to the struck quark, whereas the stretched out region in
the small-$z_{\rm jet }$ region corresponds to the proton remnant. We
do however notice that our algorithms tend to produce harder (larger
$z_{\rm jet}$) jets than Centauro in the current hemisphere.
For $R_0 = 1$, there are larger differences between the algorithms. 
The central region is populated by soft wide-angle radiation that forms stand-alone jets.
It is therefore not surprising that the $k_t$ algorithm yields more jets in this region, as it tends to cluster soft particles first. anti-$k_t$ on the other hand clusters a significant fraction of the partons in the remnant hemisphere with the current hemisphere jets, leading to fewer jets in the remnant hemisphere and harder jets in the current hemisphere. C/A and Centauro seem to be less sensitive to the increased jet radius.

\subsection{Macrojet observables at hadron level}
\label{sec:obs-macrojet}
\begin{figure}[t!]
    \centering
    \includegraphics[width=0.46\textwidth,page=1]{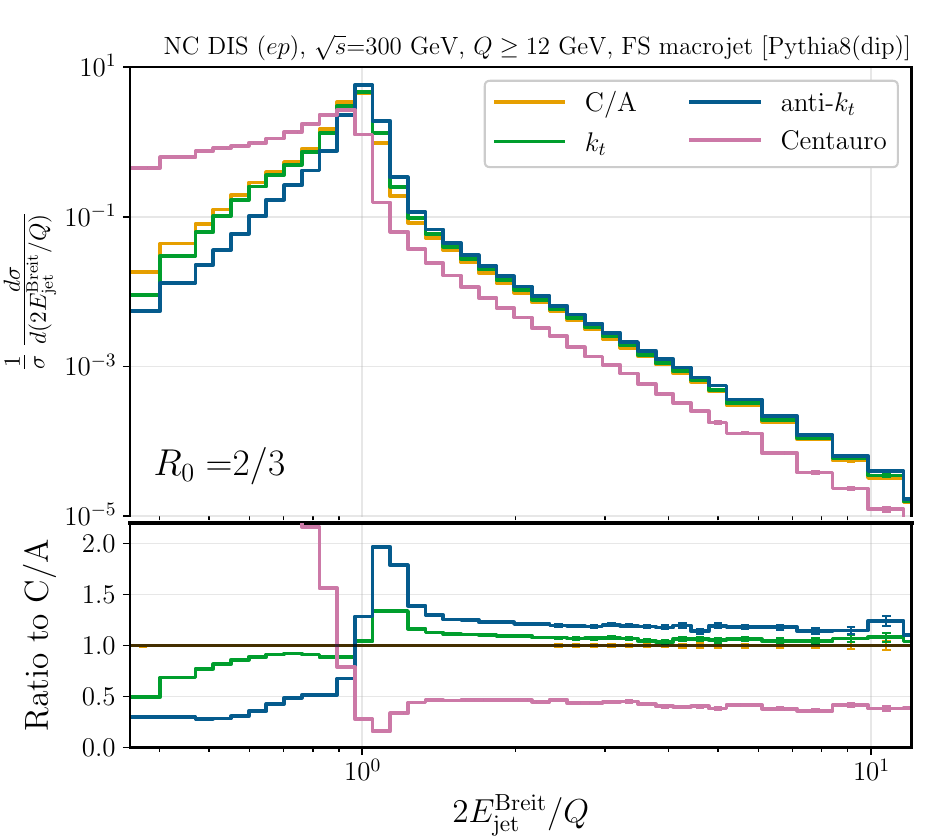}
    \includegraphics[width=0.46\textwidth,page=1]{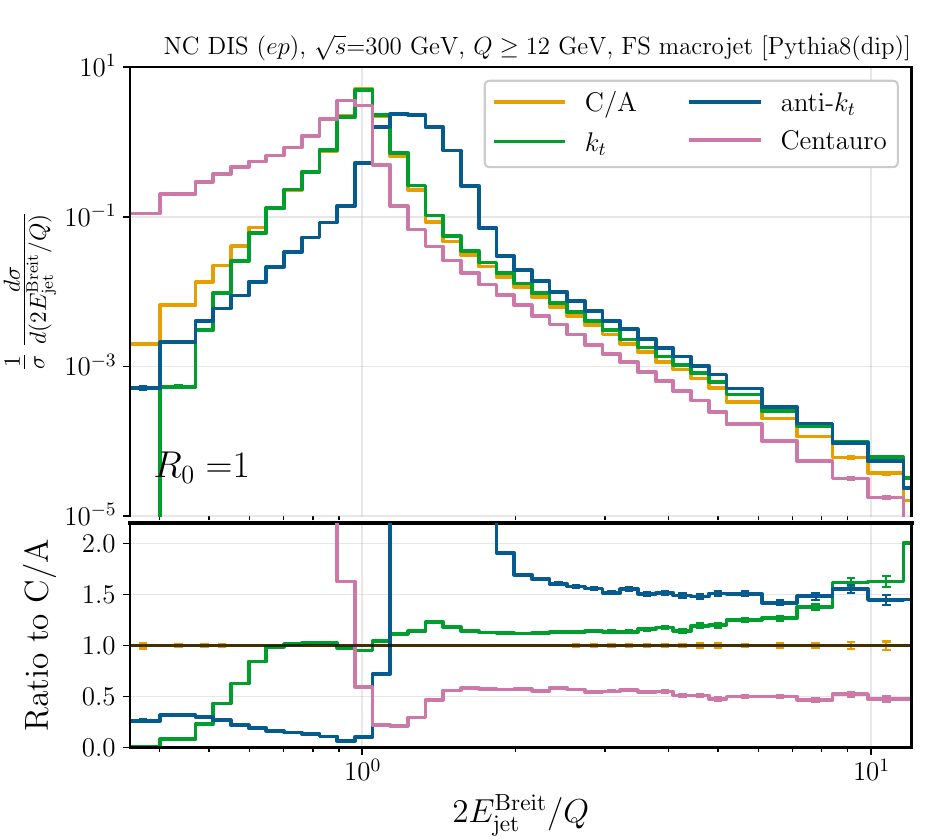}
    \caption{Normalised energy distribution of the final-state macrojet in the Breit frame for the Centauro~(pink) jet algorithm, as well as the inclusive versions of the C/A ($p=0$, orange), $k_t$ ($p=1$, green) and anti-$k_t$ ($p=-1$, blue) algorithms. We show $R_0 = 2/3$ (left) and $R_0 = 1$ (right).}
    \label{fig:E-algs}
\end{figure}

We now investigate the properties of the final-state macrojet, identified as the jet that maximises the light-cone component of the original struck quark, as explained in Sec.~\ref{sec:macrojet}.
We begin by examining the Breit-frame energy of the macrojet,
$E_{\mathrm{jet}}$, normalised to $Q/2$. At the Born-level (in the
absence of beam remnants) this variable is exactly $1$.
The normalised energy distributions of the macrojet for the
Centauro~(pink), C/A~(orange), $k_t$~(green) and anti-$k_t$~(blue)
algorithms, with $R_0=2/3$ and $R_0 = 1$, are shown in
Fig.~\ref{fig:E-algs}.
For both jet radii, the behaviour of the algorithms is very different in the soft region,
i.e.\ for $2E_{\mathrm{jet}}/Q < 1$. 
Centauro tends to produce a macrojet that is much softer than the other
three algorithms.
This can be understood by inspecting the distance measure of the Centauro algorithm
(cf.\ Eq.~\eqref{eq:cent-dij}), which effectively prevents partons from
clustering in the target hemisphere, and prevents the recoiled 
struck quark (which may lie at a more central rapidity) from clustering 
with anything that is not very close in angle.
For $R_0 = 2/3$, we observe that the slopes of all four jet algorithms 
are similar in the hard region,
i.e.\ for $2E_{\mathrm{jet}}/Q > 1$. 
In the peak region, around $2E_{\mathrm{jet}}/Q \sim 1$, we observe that
anti-$k_t$ and $k_t$ shift the peak to slightly larger values compared to C/A, 
whereas Centauro shifts it to lower values when choosing $R_0 = 2/3$. 
However, for $R_0 = 1$, we see that the behaviour in the hard region is vastly
different for all considered jet algorithms. 
In particular, all generalised-$k_t$ variants result in a peak that is  
broader for $R_0 = 1$ with respect to using $R_0 = 2/3$.
On the contrary, Centauro shows a peak that is narrower. 
We stress again that for $R_0=1$, the anti-$k_t$ variant will
have the tendency to contaminate the final-state macrojet with
beam radiation. 

\begin{figure}[t!]
	\centering
	\includegraphics[width=0.46\textwidth,page=5]{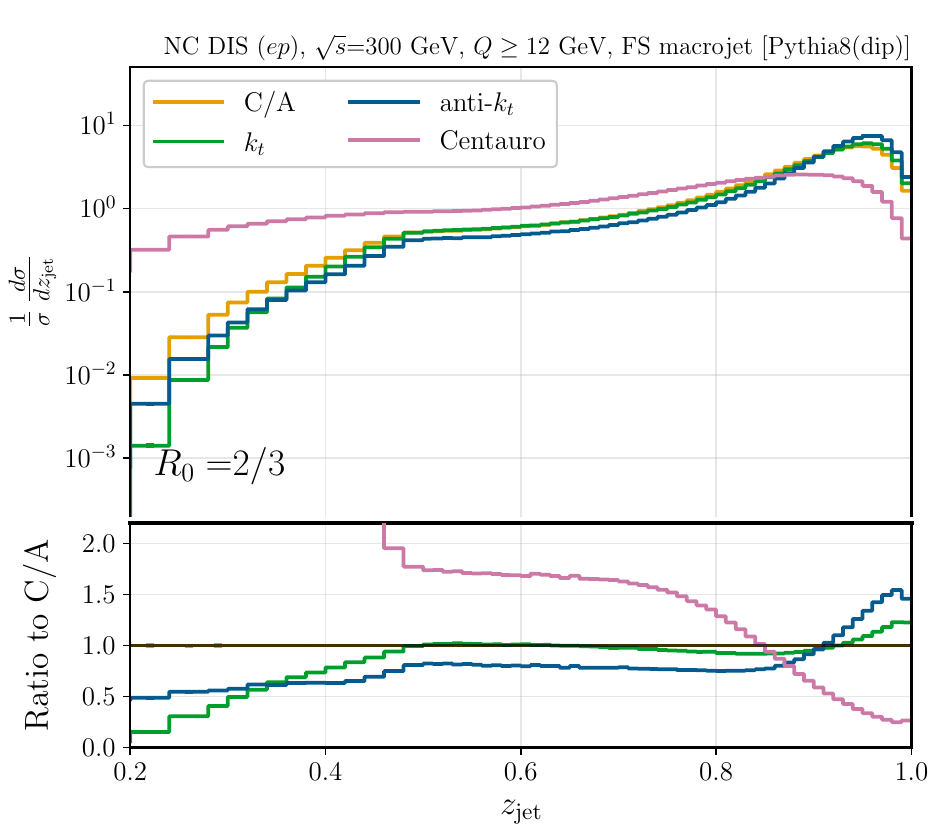}
	\includegraphics[width=0.46\textwidth,page=4]{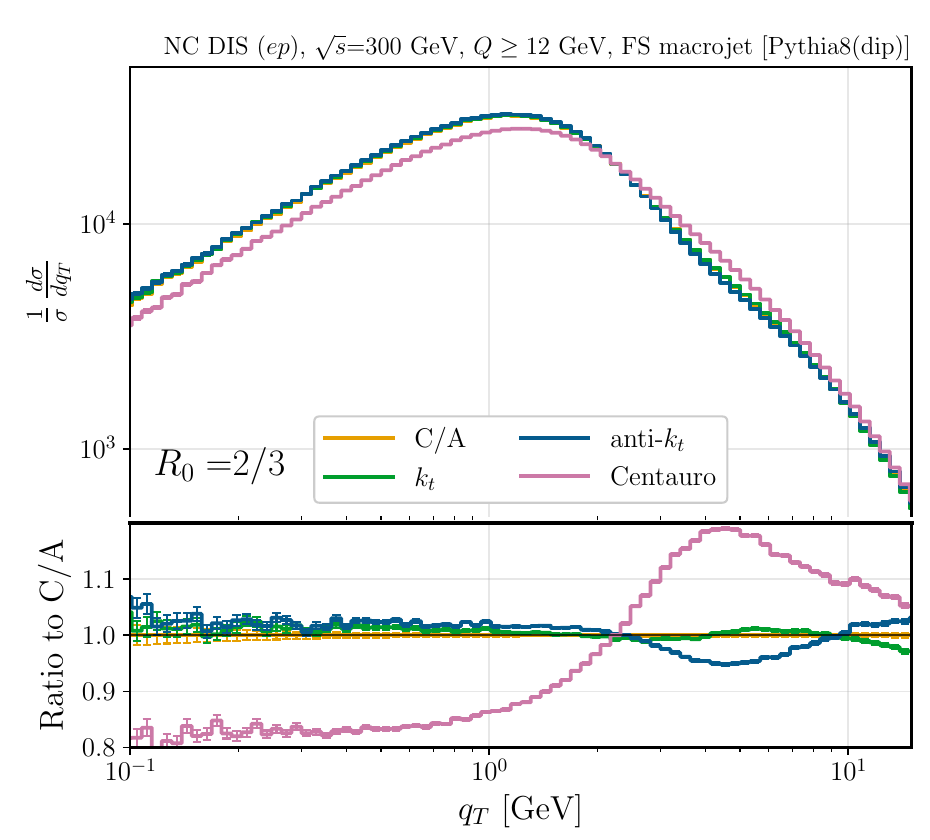}
	\caption{Similar to Fig.~\ref{fig:E-algs}, but for $z_{\rm jet}$ (left) and $q_T$ (right, see Eq.~\eqref{eq:qT}) with $R_0=2/3$.}
	\label{fig:z-qT-algs}
\end{figure}

To further investigate the behaviour of soft radiation, we examine $z_{\rm jet}$ (cf.\ Eq.~\eqref{eq:zjet}).
The resulting distributions are shown in the left panel of Fig.~\ref{fig:z-qT-algs}.
One can notice the presence of three regions:
\begin{itemize}
  \item a peak region, around $z_{\rm jet} \sim  0.95$, characterised by only soft and collinear emissions;
  \item a ``dijet" region, with $0.5 < z_{\rm jet} \lesssim 0.7$, where the final-state macrojet still carries most of the original momentum, but a non-negligible fraction has been distributed to other jets;
  \item a ``multi-jet" region, with $z_{\rm jet} <0.5$, where the struck quark lost the majority of its energy.
\end{itemize}
We notice that our generalised-$k_t$ algorithms behave quite similarly in the peak and dijet region, which terminates with a shoulder around $z_{\rm jet} =0.5$.
This similarity is expected, as these regions are dominated by configurations with at most one hard emission, and the two algorithms behave identically at $\mathcal{O}(\alpha_s)$.
Below this shoulder the distributions fall off very quickly.
The Centauro algorithm produces a quite different distribution, with
the peak region being noticeably smeared towards lower values.
The slope of the Centauro distribution agrees with the other
three jet algorithms in the dijet region. 
This is perhaps not surprising, as this is the region that is dominated by two
well-separated jets, which means the different treatment of additional soft radiation by the algorithms in their cluster sequences is less impactful.
Towards $z_{\mathrm{jet}} \to 1$, the spectrum differs significantly,
which can again be attributed to the behaviour of the $\bar{\eta}$
variable of that algorithm, which prevents some soft radiation
carrying away a fraction of the energy from being clustered into the
macrojet.
Similarly,
for $z_{\mathrm{jet}} < 0.5$ the spectrum is dominated by soft jets in
the target hemisphere. 

The last observable we study is $q_T$, which is defined as~\cite{Mulders:1995dh}
\begin{align}
\label{eq:qT}
q_T = \frac{|k_T|}{z_{\rm jet}}\,,
\end{align}
where $|k_T|$ is the modulus of the transverse momentum of the jet in the Breit frame (see Eq.~\eqref{eq:sudakov}). 
This observable is known to have little sensitivity to hadronisation, and can be used
to study the transverse momentum distribution of partons inside a proton. 
We see that all four jet algorithms show a similar slope in the very low-$q_T$ region,
but that the three generalised-$k_t$ algorithms result in 20$\%$ more events in the
region of interest, i.e.\ $q_T \lesssim Q_{\rm min} / 4 = 3 $ GeV.

\subsection{Impact of hadronisation and beam remnants}
\label{sec:hadronisation}
In this section we study the impact of hadronisation and beam remnants on the final-state macrojet. 
Given the amount of high quality DIS data that
currently exists, with more to come from the EIC, it is worth
considering if it is possible to use jets to constrain fundamental QCD
parameters, like the value of $\alpha_s$ and parameters of various
hadronisation models, as is often done with historical $e^+e^-$
collider data and LHC data. Although it is beyond the scope of this
paper to carry out a fit of such parameters, we here discuss how one
might use the jet algorithms to perform them.

\begin{figure}[t!]
  \includegraphics[width=\textwidth,page=21]{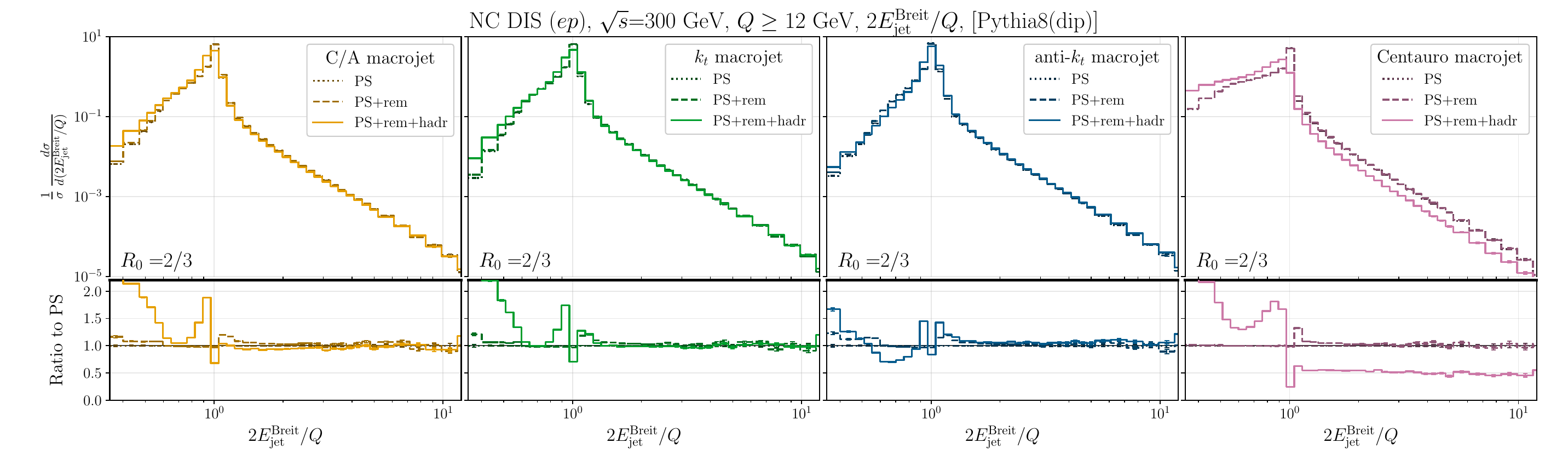}\\
  \includegraphics[width=\textwidth,page=22]{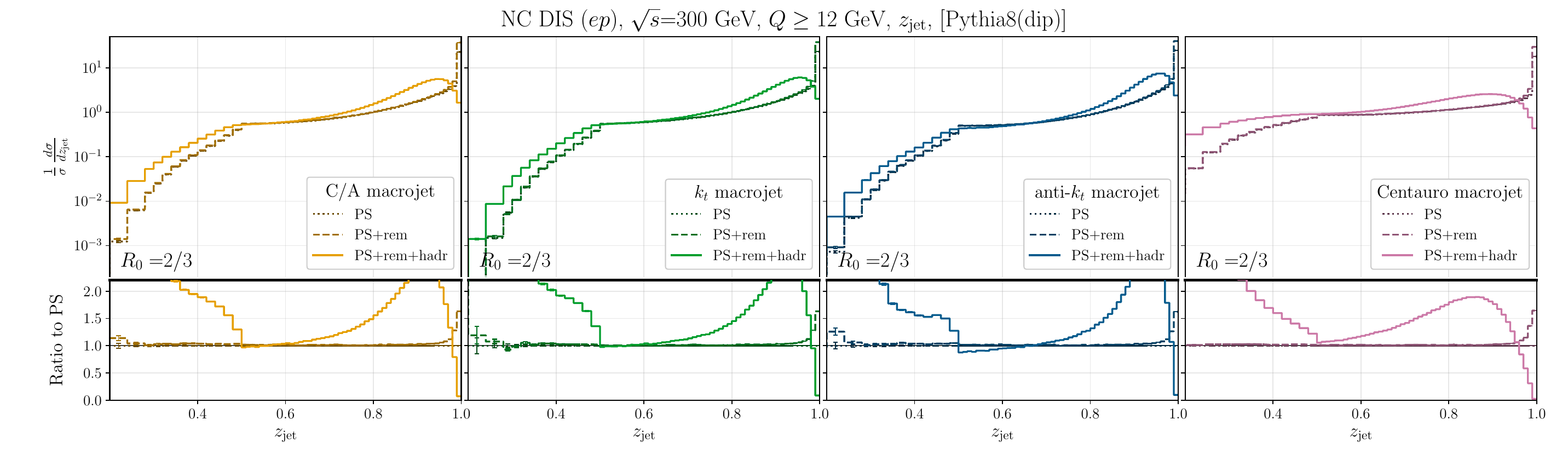}\\
  \includegraphics[width=\textwidth,page=25]{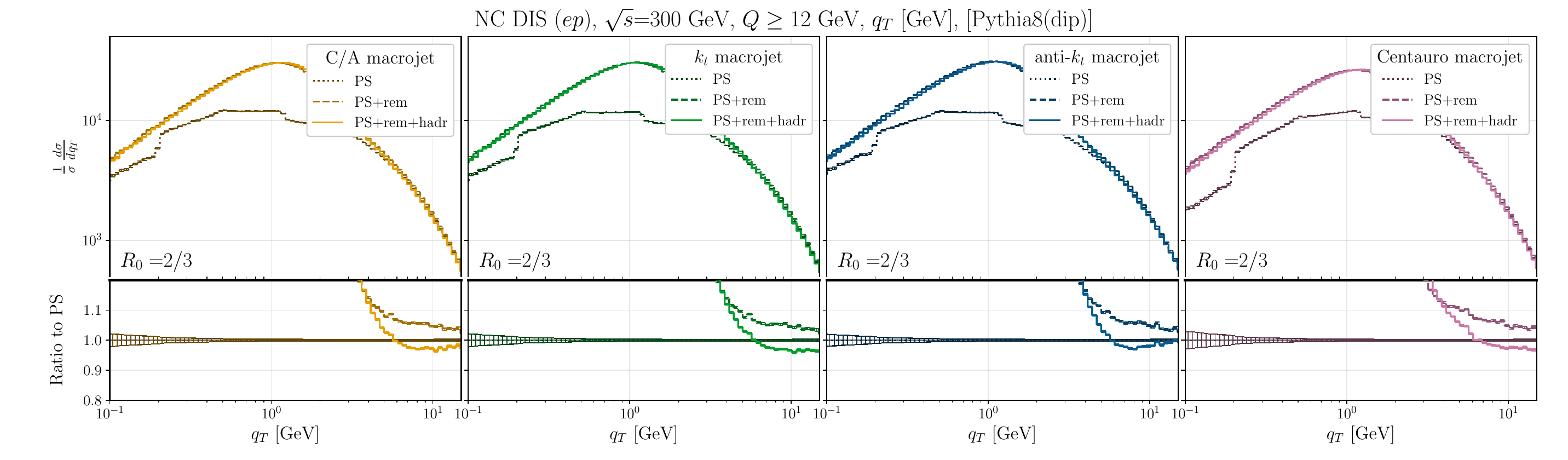}
  \caption{ Normalised energy~(first row), $z_{\rm jet}$ (second row)
    and $q_T$ (third row) distributions of the final-state macrojet in
    the Breit frame for the C/A~(first column), $k_t$~(second
    column), anti-$k_t$ (third column), and Centauro~(fourth column)
    jet algorithms, with $R_0=2/3$.
We show results at the parton shower level (labeled PS, dotted), parton shower with beam remnants (labeled PS+remn, dashed), and parton shower with beam remnants and hadronisation (labeled PS+remn+hadr, solid).}
  \label{fig:hadron}
\end{figure}
In Fig.~\ref{fig:hadron} we show the normalised energy distribution
and lightcone fraction of the final-state macrojet for the Centauro
jet algorithm, as well as the inclusive version of our generalised
$k_t$ algorithms, for $R_0=2/3$.

For each plot we show results at the parton shower level (PS, dotted), parton shower with beam remnants (PS+remn, dashed), and parton shower with beam remnants and hadronisation (PS+remn+hadr, solid). The latter distributions, which also include unstable hadrons decay, coincide with the ones shown in the previous section. 
We remind the reader, that unlike in hadron-hadron
collisions, there is no underlying event in DIS.

Starting with the normalised energy fraction we see that the beam
remnants do not contaminate the jet spectra much when compared to the parton
level results, although there is some distortion to the right of the peak. 
The hadronisation, on the other hand, has a larger effect. 
To the left of the peak, i.e.\ the soft region, we observe the expected
large distortions for all algorithms.
To the right of the peak, the normalisation for the three generalised-$k_t$ 
algorithms is surprisingly insensitive to hadronisation.
Centauro receives very large corrections of $\mathcal{O}(50\%)$ here, but the
ratio is rather flat, so we attribute the large shift to the
normalisation of the plots.

Turning now to the lightcone component $z_{\mathrm{jet}}$ we see again
that the beam remnants have very little effect on the plots, except
for $z_{\mathrm{jet}}$ close to $1$. 
Hadronisation on the other hand has once again a large effect, which induces shape differences to be
found across all the spectrum, except for just above $z_{\mathrm{jet}}=0.5$.
We also notice that for all jet algorithms, the peak at $z_{\rm jet}\approx 1$ present at parton level moves at lower values.
This observable can be used to constrain both perturbative and non-perturbative parameters.
The region $0.65<z_{\rm jet}<0.9$ can be used to fit non-perturbative
models.  
On the other hand, the region just above the shoulder, $0.5 < z_{\rm jet} < 0.65$, 
is more insensitive to hadronisation effects and could
be used to determine $\alpha_s$ by the comparison of data
to predictions.

The last observable we show is $q_T$, cf.\ Eq.~\eqref{eq:qT}. 
Contrary to the previously discussed observables, 
here we see that the hadronisation has relatively little impact
compared to the addition of beam remnants. 
As said before, the transverse momentum distribution of
partons in the proton may be studied using this observable,
as can be clearly seen from these results. 
For the generalised-$k_t$ algorithms we see again that more
events enter the region of interest ($q_T \lesssim 3$) compared
to the Centauro algorithm. 
Finally we note that the parton-shower only results are not fully
physical, which shows itself through kinks at low $q_T$, which are
associated with various shower cut offs.

\section{Conclusion}
\label{sec:concl}
In this paper we have presented a generalisation of the
angular-ordered jet algorithm which was first presented in
Ref.~\cite{vanBeekveld:2023chs}.  The generalisation follows the idea
of introducing an exponent to control how much to weight the energy of
each pseudojet in the clustering, providing a tunable $p$-family
familiar from similar lepton- and hadron-collider algorithms. The
algorithm is very simple, and gives jets that are naturally separated
in the current and target hemispheres.  We also compared the
angular-ordered, $k_t$ and anti-$k_t$ variants of our new inclusive
algorithm to the recently-proposed Centauro jet algorithm. One
application that we have in mind is to use the generalised $k_t$
algorithms to re-analyse HERA data, with the aim of providing
complementary constraints on \(\alpha_s\) from jets, and of
constraining hadronisation and intrinsic $k_t$ parameters in event
generators. Whereas an $\alpha_s$ extraction is unlikely to be
competitive with other extractions on its own, we think there is value
in studying non-perturbative physics in DIS data. This is in part due
to the fact that DIS allows for studying jets in bins of $Q$. This
could provide a natural handle on the energy dependence of
non-perturbative parameters. We leave this to future studies.
The algorithm has been implemented as the \href{https://repo.hepforge.org/source/fastjetsvn/browse/contrib/contribs/DISGenkt}{\texttt{DISGenkt}} plugin to
\textsc{FastJet} as part of \texttt{fjcontrib}.

\section*{Acknowledgments}
We thank Gavin Salam and Gregory Soyez for useful discussions regarding
differences between the Centauro and the generalised $k_t$ jet
algorithms.  In particular, we thank Gavin Salam for useful
discussions on the formulation of the anti-$k_t$ variant presented
here.  This research was supported by the Italian Ministry of
Universities and Research (MUR) under the FIS grant (CUP:
D53C24005480001, FLAME) (SFR) and by the Dutch Research Council (NWO)
under project number VI.Veni.232.190 (MvB). DP was supported by the
Science and Technology Facilities Council (STFC) under the Studentship
Grant ST/X508822/1 and is grateful for the hospitality of the CERN TH
Department while this research was carried out. AK acknowledges
funding from a Royal Society Research Professorship (grant
RP$\backslash$R$\backslash$231001), and acknowledges the CERN TH
Department for hospitality while this research was being carried out.

\bibliography{disjet.bib}

\nolinenumbers

\end{document}